\documentclass[11pt]{article}

\usepackage[margin=1in]{geometry}
\usepackage[utf8]{inputenc}
\usepackage[T1]{fontenc}
\usepackage{lmodern}
\usepackage{amsmath,amssymb}
\usepackage{graphicx}
\usepackage{booktabs}
\usepackage{array}
\usepackage{xcolor}
\usepackage{tikz}
\usetikzlibrary{arrows.meta,positioning,shapes.geometric,calc,fit}
\usepackage{pgfplots}
\pgfplotsset{compat=1.18}
\usepackage{microtype}
\usepackage[colorlinks=true,linkcolor=blue,citecolor=blue,urlcolor=blue]{hyperref}
\hypersetup{
  pdftitle={Machine learning for rarefied gas transport in vacuum and micro/nano systems: promise, pitfalls, and a verification agenda},
  pdfauthor={Ehsan Roohi},
  pdfsubject={Rarefied gas dynamics and machine learning},
  pdfkeywords={Rarefied gas dynamics, DSMC, machine learning, physics-informed neural networks, operator learning, vacuum technology}
}

\title{Machine learning for rarefied gas transport in vacuum and micro/nano systems: promise, pitfalls, and a verification agenda}
\author{Ehsan Roohi\\
Department of Mechanical and Industrial Engineering\\
University of Massachusetts Amherst\\
Amherst, MA 01003, USA\\
\href{mailto:roohie@umass.edu}{roohie@umass.edu}}
\date{June 2026}

\begin{document}

\maketitle

\begin{abstract}
Machine learning is beginning to influence rarefied-gas modeling at
multiple levels, including equation-solving, operator learning, learned
collision physics, moment closures, direct simulation Monte Carlo (DSMC) field surrogates, and
gas--surface models. This Perspective argues that the central challenge is
not demonstration-level success, but trustworthy use under realistic
deployment conditions: multiregime Knudsen behavior, stochastic DSMC
labels, sharp nonequilibrium structures, uncertain gas--surface
interaction, and scarce direct experimental anchors. I classify the main
method families by what is learned, distinguish soft physics penalties
from structure-preserving designs, and propose evaluation standards based
on extrapolation tests, noise-aware metrics, end-to-end cost accounting,
and a three-level validation hierarchy. Most current evidence is
solver-facing: it demonstrates surrogate fidelity to a teacher solver more
often than direct physical fidelity to experiment. The aim is not to
dismiss ML for rarefied and vacuum-related gas transport, but to separate
what is already credible from what remains provisional, and to define a
reporting standard that makes future claims auditable.
\end{abstract}

\noindent\textbf{Keywords:} Rarefied gas dynamics; DSMC; machine learning; physics-informed neural networks; operator learning; vacuum technology; Knudsen number; gas--surface interaction

\vspace{1em}
\section{Introduction}

Rarefied gas transport sits at the core of vacuum science and technology.
The design of ultra-high-vacuum beam lines, cryopumps, getters, and gauges;
the operation of micro-electro-mechanical systems (MEMS), Knudsen pumps,
and micro gas chromatographs; the prediction of satellite aerodynamics in
very low Earth orbit; and the aerothermodynamics of re-entry vehicles all
involve flows in which the Knudsen number $\mathrm{Kn}=\lambda/L$ is no
longer vanishingly small, so that the Navier--Stokes--Fourier (NSF)
description loses validity and kinetic theory takes over
\cite{bird1994,cercignani1988,sharipov2016,akhlaghi2023physrep}. The
community's reference tools are the direct simulation Monte Carlo (DSMC)
method \cite{bird1994,roohi2016physrep,roohibook2025}, deterministic
discrete-velocity and unified gas-kinetic solvers
\cite{mieussens2000,xu2010ugks}, test-particle Monte Carlo codes such as
Molflow+ for free-molecular vacuum-system design \cite{kersevan2009}, and,
for moderately rarefied conditions, extended-hydrodynamic moment closures.
These tools are accurate, but they are expensive: a single
three-dimensional DSMC simulation of a transitional-regime device can
consume thousands of CPU-hours even on optimized, massively parallel
implementations \cite{plimpton2019,scanlon2010}, and decades of
algorithmic refinement of collision schemes have improved constants rather
than removed the fundamental cost scaling \cite{roohi2016physrep,%
shojasani2026vacuum}.

The cost problem is most acute not for one-off simulations but for
\emph{many-query} workflows. Parametric studies of a vacuum component
across geometry and operating pressure, design optimization of a Knudsen
pump stage, uncertainty quantification with respect to accommodation
coefficients, real-time model-based control, and inverse estimation of
boundary parameters from sparse gauge data may each require $10^2$--$10^5$
forward solves. It is precisely here that machine learning (ML) entered
the field. Since roughly 2019, physics-informed neural networks (PINNs)
have been applied to Boltzmann-model equations \cite{lou2021,%
deflorio2021}, neural operators have been trained to map geometry and
flow parameters to entire rarefied flow fields \cite{lu2021deeponet,%
peyvan2026,roohi2026mnf}, neural networks have been embedded inside
kinetic solvers as surrogates of the collision integral
\cite{xiao2021,xiao2023relaxnet,roohi2026collision} or of moment closures
and constitutive relations \cite{han2019,sadr2020,schotthofer2022,%
garg2024ncf,garg2025diatomic}, regression and surrogate models have
been used to couple continuum and particle descriptions with DSMC-derived
corrections \cite{sadr2021coupling,tatsios2025mms,chinnappan2025bayesian},
and data-driven scattering kernels have been constructed from
molecular-dynamics data \cite{andric2019,roohi2026kernels}. The literature is growing quickly,
but it is fragmented across the kinetic-theory, scientific-ML, aerospace,
and vacuum-technology communities, and the literature's evaluation
practices remain uneven relative to the scope of the claims being made.

This article is a critical Perspective rather than a neutral survey. The
method-family classification in Section~\ref{sec:landscape} is meant to
orient readers before the critique intensifies; it is not an exhaustive
bibliography. I use representative, method-defining papers and recent
application studies to support field-level claims. Mature peer-reviewed
contributions are used as the main evidence for settled methodological
points. Recent 2025--2026 papers, including several from my own group,
are used as case studies and, where they are preprints, should be read as
provisional evidence rather than field consensus. This source-status
convention is important because some of the sharpest examples in this
fast-moving literature are too recent to carry the evidentiary weight of
canonical peer-reviewed benchmarks.

The rest of the article follows the evidence standard it recommends.
Section~\ref{sec:landscape} classifies the main method families by what is
learned and by what can be guaranteed. Section~\ref{sec:why} explains why
rarefied gas transport is a particularly stringent test for ML rather
than just another smooth-PDE benchmark. Section~\ref{sec:constraints}
examines what ``physics-informed'' does and does not guarantee, including
the identifiability limits that follow from training on macroscopic data.
Section~\ref{sec:experiment} presents a three-level validation hierarchy
and discusses the structural obstacles to direct experimental
verification. Section~\ref{sec:challenges} turns these critiques into a
reporting checklist, and Section~\ref{sec:outlook} closes with a roadmap
framed as falsifiable milestones rather than broad verdicts.

Two framing points should be kept explicit throughout. First, the question
is not whether ML can produce attractive demonstrations; by now it can.
The question is whether a specific ML model is \emph{trustworthy under the
conditions in which it would actually be used}---outside the training
range, with noisy kinetic data, with uncertain boundary conditions, and
with no reference solution available to check against. Second,
acceleration claims must be evaluated end to end. A surrogate that is four
orders of magnitude faster than DSMC per query but requires hundreds of
DSMC solutions for training has a break-even point, and whether a given
application ever reaches it is an engineering question that should be
reported rather than assumed.
\section{The current landscape: a critical taxonomy}\label{sec:landscape}

Six method families dominate the literature. Table~\ref{tab:taxonomy}
summarizes them; the subsections give a compressed critical reading of
each. The classification follows \emph{what is learned}, since that
determines which physical guarantees are even possible.

\begin{table*}[!tp]
\centering
\caption{Method families for ML in rarefied gas transport. ``Guarantees''
refers to properties enforced by construction (hard), as opposed to
penalized in the loss (soft) or merely observed a posteriori.
``Validation status'' reflects the dominant practice in the literature,
not the best single paper.}
\label{tab:taxonomy}
\small
\setlength{\tabcolsep}{4pt}
\begin{tabular}{>{\raggedright\arraybackslash}p{2.5cm}>{\raggedright\arraybackslash}p{2.8cm}>{\raggedright\arraybackslash}p{2.7cm}>{\raggedright\arraybackslash}p{1.5cm}>{\raggedright\arraybackslash}p{2.7cm}}
\toprule
Family & What is learned & Typical guarantees & Data appetite &
Validation status \\
\midrule
PINN kinetic solvers \cite{lou2021,deflorio2021} &
$f$ or its moments for one configuration &
Soft PDE/BC residuals; hard BCs only in special constructions &
Low (equation driven) &
Verified against deterministic solvers on canonical 1D and 2D cases \\
\addlinespace
Operator learning \cite{lu2021deeponet,peyvan2026,roohi2026mnf} &
Parameter or geometry $\to$ field maps &
None by default; weighting heuristics for shocks &
High ($10^2$-- $10^3$ solves) &
Solver-vs-surrogate only; splits often interpolative\strut \\
\addlinespace
Neural collision operators \cite{xiao2023relaxnet,roohi2026collision} &
Collision integral or collision-event outcomes &
Conservation, positivity, H-theorem hard in best designs &
Moderate &
In-the-loop tests vs.\ DSMC and Boltzmann benchmarks \\
\addlinespace
Learned moment closures and constitutive corrections \cite{han2019,sadr2020,schotthofer2022,garg2024ncf,garg2025diatomic} &
Closure relations, entropy maps, or DSMC-informed constitutive terms &
Hyperbolicity, realizability, entropy dissipation achievable by design &
Moderate &
Canonical kinetic benchmarks; growing shock/shear-flow ML tests \\
\addlinespace
DSMC field/correction surrogates \cite{roohi2026ast,roohi2026nnrgd,tatsios2025mms,chinnappan2025bayesian} &
End-to-end regression of DSMC fields or DSMC-to-continuum corrections &
None in pure field maps; conservation retained in correction-based hybrids &
High &
Solver-vs-surrogate; noise floor partly addressed in hybrid-correction work \\
\addlinespace
Data-driven GSI kernels \cite{andric2019,roohi2026kernels} &
Scattering kernels from MD/beam data &
Reciprocity and normalization enforceable; transport preservation in recent work &
MD-limited &
Molecular-beam comparisons in isolated cases; flow-level validation rare \\
\bottomrule
\end{tabular}
\end{table*}
\subsection{PINNs on kinetic model equations}

PINNs minimize the residual of a governing equation (plus boundary and
data terms) over a neural ansatz \cite{raissi2019,karniadakis2021}.
Applied to the Boltzmann--BGK equation, they have solved forward and
inverse problems for Couette and cavity flows from the continuum limit to
$\mathrm{Kn}=10$ \cite{lou2021}, and, combined with functional
interpolation that imposes boundary conditions exactly, have reproduced
classical linear-kinetic benchmarks such as thermal creep between plates
with high accuracy \cite{deflorio2021}. Two honest readings coexist. The
favorable one: PINNs require no labeled data, handle inverse problems
naturally (e.g., reconstructing a field from scattered interior
measurements without knowing the boundary condition \cite{lou2021}), and
are therefore attractive precisely where DSMC data generation is the
bottleneck. The unfavorable one: for \emph{forward} problems on model
equations, mature deterministic discrete-velocity solvers
\cite{mieussens2000,xu2010ugks} are faster, more accurate, and carry
convergence theory; published PINN results in this area are validated
against those very solvers, which concedes the point. The defensible
niche for kinetic PINNs is therefore inverse and data-assimilation
settings, plus configurations with parametric embedding---not routine
forward solution. Add to this the well-known PINN pathologies (stiff
multi-scale training, spectral bias against the sharp structures discussed in
Section~\ref{sec:why}, sensitivity to loss weights), which are amplified rather
than diminished by kinetic equations, and the appropriate posture is
selective use, not general adoption.
\subsection{Operator learning for parametric and geometric maps}

Neural operators---DeepONet \cite{lu2021deeponet}, Fourier neural
operators \cite{li2021fno}, and their descendants---learn maps between
function spaces and are the natural formalism for many-query problems:
one trained model answers a continuum of parameter or geometry queries.
In high-speed and rarefied applications they have been pushed
furthest by geometry-conditioned variants such as Fusion-DeepONet for
hypersonic and supersonic shape-dependent flows \cite{peyvan2026}, and by
hybrid frameworks coupling operator networks to chemistry surrogates
behind normal shocks \cite{mao2021}. In the author group's work, the same
machinery has been adapted to rarefied internal flows: shock-aware,
physics-guided Fusion-DeepONet for micro-nozzle fields
\cite{roohi2025fusion}, and DeepONet with a zonal loss that partitions a
separated micro-step flow into physically distinct regions with
region-specific penalties \cite{roohi2026mnf}.

Three criticisms apply to this family as practiced, including by us.
First, the evaluation protocols are dominated by random splits over dense
parameter sweeps; since neighboring DSMC cases are strongly correlated,
such tests measure interpolation between near-duplicates and say almost
nothing about the design-exploration use case that motivates the work
(Section~\ref{sec:challenges}). Second, baselines are weak or absent:
when the underlying parameter-to-field map is smooth and effectively
low-rank---which is common in micro-flow sweeps over inlet pressure or
wall temperature---linear reduced-order models or Gaussian-process
regression on POD coefficients can match deep operators at a fraction of
the cost, and a deep architecture should be required to beat that
baseline, not merely to beat ``no surrogate''. Third, the offline cost is
under-reported: a training corpus of several hundred DSMC solutions is
itself a major computational campaign, and the honest figure of merit is
the break-even query count, not the per-query speed-up.
\subsection{Neural surrogates inside the kinetic solver: collision
operators and collision events}

A different philosophy keeps the kinetic solver and replaces its most
expensive kernel. For deterministic solvers, the target is the fivefold
collision integral: Xiao and Frank's RelaxNet provides a
structure-preserving surrogate that provably maintains positivity,
conservation, the H-theorem, and the correct hydrodynamic limit
\cite{xiao2023relaxnet}, building on earlier neural acceleration of
Boltzmann solution operators \cite{xiao2021}. For particle methods, the
target is the per-event physics: in our recent work, physics-constrained
neural collision operators emulate variable-hard-sphere collisions and
predict ab initio--informed scattering angles inside DSMC
\cite{roohi2026collision}, and deep-learning--accelerated Lennard-Jones
collision modeling extends DSMC toward cryogenic and high-temperature
regimes where simple phenomenological potentials degrade
\cite{shojasani2026lj}.

This family deserves, in my assessment, the most optimism relative to its
promotion, for a structural reason: the surrounding solver continues to
enforce transport, boundary conditions, and (in DSMC) exact
conservation at the pair level if the surrogate is constrained to supply
momentum- and energy-conserving post-collision states. The error
committed by the network is local and is mediated by the surrounding
physically constrained algorithm, instead of accumulating in an
unconstrained end-to-end map. The criticisms are correspondingly
specific rather than existential: (a) speed-ups are bounded by Amdahl's
law---if collisions are 40--70\% of runtime, no collision surrogate
delivers orders of magnitude end to end, and claims should be stated
against total wall-clock time on stated hardware; (b) out-of-distribution
collision energies (strong shocks, plume far fields) will be visited by
the solver whether or not they were sampled in training, so the sampling
strategy is a correctness issue, not a tuning detail---the
closure-hierarchy sampling of \cite{xiao2023relaxnet} is a model
response; (c) for DSMC specifically, the surrogate must not distort the
collision-rate statistics that decades of collision-scheme analysis have
been devoted to getting right \cite{roohi2016physrep,shojasani2026vacuum},
which argues for verifying learned-collision DSMC against the same
transport-coefficient and relaxation benchmarks used to certify
conventional schemes.
\subsection{Learned moment closures}

Moment methods trade the velocity dimension for a closure problem, and
the closure is a map---an ideal target for regression. The benchmark
contributions are instructive precisely because they took structure
seriously: Han et al.\ learned hydrodynamic models from kinetic data
with uniform accuracy across regimes by learning a generalized set of
moments first \cite{han2019}; Sadr, Torrilhon and Gorji replaced the
expensive maximum-entropy optimization with Gaussian-process regression
of the Lagrange multipliers, retaining the realizability advantages of
the entropy family \cite{sadr2020}; Schotthöfer et al.\ embedded
convexity into the network so that entropy dissipation and hyperbolicity
of the closed system survive the approximation \cite{schotthofer2022}.
On the particle side, our group is pursuing the same logic for
Fokker--Planck--based solvers, substituting the moment-closure evaluation
with a GPU-native network to remove the dominant cost of the cubic
Fokker--Planck scheme. Related in spirit, symbolic
approaches have recovered constitutive equations directly from molecular
simulation data \cite{zhang2020jfm,rudy2017,brunton2016}, which offers
interpretability that black-box closures lack.

A parallel and practically important line learns the correction or
constitutive information that a continuum solver needs, rather than
learning the entire flow field. Data-driven maximum-entropy distributions
have been used to couple kinetic and continuum descriptions
\cite{sadr2021coupling}; neural finite-volume--DSMC frameworks have
learned nonlinear constitutive relations from DSMC data for monatomic and
diatomic shock structures and for velocity-shear dominated flows
\cite{garg2024ncf,garg2025diatomic,garg2026topology}; and surrogate or
Bayesian-regression models have been used to infer DSMC-derived stress,
slip, and boundary corrections for low-speed rarefied flows
\cite{tatsios2025mms,chinnappan2025bayesian}. These methods are not
mere black-box field regressors: by embedding the learned information as
corrections inside conservation-law solvers, they occupy a useful middle
ground between classical hybrid continuum--particle methods and fully
learned end-to-end surrogates.

The critical issues here are sharply defined. Realizability: a closure
that leaves the realizable moment set produces unphysical states and
crashes solvers; only architectures that respect the set by construction
\cite{sadr2020,schotthofer2022} avoid this failure mode, and soft
penalties demonstrably do not. Hyperbolicity: a learned closure that
breaks hyperbolicity yields ill-posed initial-value problems whose
numerical symptoms (grid-dependent oscillations) are easily misdiagnosed
as discretization error. And regime honesty: closures trained on
near-equilibrium data will be smooth and numerically stable but physically inaccurate in
strongly non-equilibrium regions; the failure may not be obvious, which
makes it consequential.
\subsection{End-to-end DSMC field surrogates}

The most popular family, because it is the easiest to execute: run a DSMC
parameter sweep, train a network from parameters (and coordinates) to
fields, report errors of a few percent and speed-ups of $10^2$--$10^5$.
Our own contributions include deep-network surrogates of DSMC solutions
for micro/nano and hypersonic configurations \cite{roohi2026ast} and
physics-enforced variants spanning relaxation problems, polyatomic shock
waves, and hypersonic cylinder flow \cite{roohi2026nnrgd,%
roohi2025enforced}. I will not rehearse the successes, which are real
within their evaluation envelope; the purpose of this subsection is to
state the envelope.

Such surrogates are best understood as \emph{compressed interpolants of a
specific solver, under a specific physical sub-model set, over a specific
parameter box}. Every term in that sentence is a limitation. ``Specific
solver'': the surrogate inherits the model-form error of the training
code (collision model, internal-energy treatment, chemistry) and cannot
be more right than its teacher---a point that becomes central in
Section~\ref{sec:experiment}. ``Sub-model set'': change the accommodation
coefficient and the training corpus is silently invalid. ``Parameter
box'': outside it, smooth extrapolation is a property of the activation
functions, not of gas dynamics. Within the envelope, the remaining
hazards are statistical: noise-floor-blind metrics (Section~\ref{sec:why},
item~iii) and leaky splits (Section~\ref{sec:challenges}). Uncertainty
quantification via Monte Carlo dropout or ensembles \cite{gal2016,%
psaros2023} is increasingly bolted on, but it must be said plainly that
these techniques quantify the model's self-declared uncertainty, which is
poorly calibrated exactly where it matters---out of distribution---and is
no substitute for physical audits.
\subsection{Data-driven gas--surface interaction}

GSI is simultaneously the most consequential and least developed target
for ML in this field. Andric, Meyer and Jenny demonstrated the canonical
pipeline: molecular-dynamics scattering data compressed into a
nonparametric, resampleable kernel that respects reciprocity and
reproduces molecular-beam observations better than classical parametric
kernels \cite{andric2019}. Our recent contribution constructs neural ab
initio scattering kernels for binary mixtures with transport-coefficient
preservation built into the training objective, so that the learned
kernel cannot improve microscopic fidelity at the price of corrupting
viscosity, diffusion, and thermal-transpiration behavior
\cite{roohi2026kernels}.

The criticisms here are mostly about the data, not the learning. MD
training data inherit the uncertainty of the interatomic potential and of
the idealized (clean, crystalline) surface; real engineering surfaces in
vacuum systems are rough, oxidized, and contaminated, with effective
accommodation drifting over operating history---which is exactly what the
sparse experimental record shows \cite{trott2011,sharipovmoldover2016,%
sharipov2011data}. A learned kernel that is exquisitely faithful to MD of
a pristine surface may therefore be \emph{less} predictive for a real
chamber wall than a crude Maxwell kernel with an empirically fitted
accommodation coefficient. The constructive path is uncomfortable but
clear: flow-level validation of learned kernels against measured
quantities that are GSI-dominated (microchannel mass flow,
thermal-transpiration ratios, free-molecular conductances), and
ML-assisted \emph{inference} of GSI parameters from such experiments with
honest uncertainty---for which the identifiability questions of the next
section are decisive.
\subsection{The free-molecular limit: the regime the ML literature
ignores}

A final observation on the taxonomy is conspicuous by the absence it
documents: nearly all of the work surveyed above targets the transition
regime, $\mathrm{Kn} \sim 0.01$--$10$, where the collision integral is
the computational bottleneck and therefore the natural object to
approximate or accelerate. Yet a large fraction of vacuum engineering
operates in the free-molecular limit, $\mathrm{Kn} \gg 10$, where
collisions are irrelevant and the entire physics resides in geometry and
gas--surface interaction: transmission probabilities, conductances of
ducts, elbows, and screens, and the angular redistribution encoded in the
Knudsen cosine law or its non-diffuse corrections. Here the established
tools---analytical view-factor methods and test-particle Monte Carlo
codes such as Molflow+ \cite{kersevan2009}---are already fast per
evaluation, so naive ``ML acceleration'' of a single solve has little to
offer. What is expensive is the \emph{outer loop}: conductance and
pumping-speed evaluation repeated over thousands of candidate geometries
in the design of accelerator beam pipes, cryopumping arrays, and
semiconductor process chambers, with surface properties (sticking and
accommodation coefficients, outgassing rates) that are themselves
uncertain and drifting.

Most recent ML work in this area targets transitional flows, where
collision modeling is expensive and method development is mathematically
fashionable. Yet many practical vacuum-design tasks lie in the
free-molecular regime, where single evaluations are already manageable
and the real cost lies in repeated geometry and surface-parameter studies.
That regime may be a better near-term target for experimentally anchored
surrogates, because the oracle is controllable, the physics is simpler,
and conductance or pump-down observables are measurable. In the targeted
literature reviewed for this Perspective, I did not identify a published
geometry-conditioned neural surrogate for free-molecular conductance that
was validated against measurement. I state that as a scoped gap rather
than a universal negative. For the vacuum-science readership, a
measured-conductance-validated, geometry-conditioned surrogate for
free-molecular components would be among the most valuable and most
achievable contributions currently open (Section~\ref{sec:outlook}),
precisely because the free-molecular limit removes the collision-modeling
uncertainty and isolates the two ingredients vacuum practice cares about:
geometry and GSI.
\section{Why rarefied gas transport is a stringent test for machine
learning}\label{sec:why}

It is tempting to treat rarefied gas dynamics as just another PDE-driven
field to which the standard scientific-ML toolbox can be ported. Five
structural features argue against this view.

\paragraph{(i) The state space is a distribution function}
The fundamental unknown is the velocity distribution function
$f(\mathbf{x},\mathbf{v},t)$ on a phase space of up to six dimensions plus
time, governed by the Boltzmann equation with its fivefold collision
integral \cite{cercignani1988}. Any ML model that operates only on
macroscopic fields (density, velocity, temperature, stress, heat flux)
has implicitly performed a drastic, lossy projection. In near-equilibrium
conditions this projection is benign; in transitional and free-molecular
conditions it discards exactly the information---non-Maxwellian tails,
bimodality inside shocks and Knudsen layers---that distinguishes kinetic
from continuum physics. The consequences for identifiability are taken up
in Section~\ref{sec:constraints}.

\paragraph{(ii) Multi-regime behavior over decades of Knudsen number}
Vacuum and micro/nano applications routinely span
$\mathrm{Kn}\sim10^{-3}$--$10^{2}$ within a single device or mission
profile: a micro-nozzle expands from near-continuum throat conditions
into a free-molecular plume; a vacuum chamber during pump-down traverses
the entire transition regime. Plain neural networks possess no mechanism
that guarantees correct asymptotic behavior in either limit, and models
trained in one regime extrapolate poorly into another. The
asymptotic-preserving neural network program of Jin and co-workers
\cite{jin2023apnn,jin2024apnn} demonstrates both that this failure is
real and that it can be mitigated by construction rather than by data
volume; analogous discipline is still rare in the rarefied-flow ML
literature.

\paragraph{(iii) The reference data are stochastic}
DSMC, the dominant data source for supervised learning in this field, is
a Monte Carlo method. Its field estimates carry statistical errors with
well-characterized scaling \cite{hadjiconstantinou2003}; DSMC convergence
and solution-verification criteria have also been analyzed through
conservation-based viewpoints \cite{karchani2015,myong2019verification}.
For low-speed micro flows the signal-to-noise ratio in velocity is
notoriously poor unless enormous sample counts are used. Because DSMC is a stochastic
teacher, surrogate accuracy should be interpreted relative to label
uncertainty rather than as an isolated percentage. In practice, this
means reporting per-field noise estimates, comparing surrogate error with
independent-run solver discrepancy, and making clear whether the
surrogate is outperforming the teacher's sampling noise or merely
tracking one realization of it. Errors below the estimated label-noise
level should trigger a replicate-based audit before being interpreted as
physical accuracy. The noise is also spatially correlated through the
sampling procedure, violating the independence assumptions behind naive
train/test error estimates. Remarkably few papers in this area report the
noise level of their training fields next to their test metrics; this
omission can materially distort how reported accuracy should be
interpreted.

\paragraph{(iv) Boundaries dominate, and boundaries are uncertain}
In internal vacuum and micro/nano flows, gas--surface interaction (GSI)
is frequently the single most influential physical sub-model: mass flow
through microchannels, thermal-transpiration pumping, radiometric forces,
and free-molecular conductances all depend sensitively on accommodation
coefficients \cite{sharipov2011data,akhlaghi2023physrep}. Yet measured
accommodation coefficients vary with gas, surface material, roughness,
temperature, and contamination history, and the experimental record,
while careful, is sparse \cite{trott2011,sharipovmoldover2016}. An ML
surrogate trained on DSMC data generated with full accommodation inherits
that assumption silently; its reported accuracy is conditional on a
boundary model whose own uncertainty is rarely propagated. The same
issue undermines ML-based inverse inference of GSI parameters from flow
data, as discussed in Sections~\ref{sec:constraints}
and~\ref{sec:experiment}.

\paragraph{(v) Sharp structures and similarity breakdown}
Shock waves, Knudsen layers of thickness comparable to the mean free
path, and expansion-fan/plume structures concentrate the approximation
difficulty into thin regions that standard $L_2$ training losses
systematically underweight. For wall-bounded flows, the same Knudsen-layer
physics has motivated higher-order continuum corrections and second-order
constitutive/slip-jump models \cite{lockerby2005knudsen,myong2016knudsen}. In hypersonic rarefied conditions the
difficulty compounds: shock stand-off and structure cease to obey
continuum similarity correlations as rarefaction increases, so a model
trained on near-continuum cases has no scaling law to fall back on.
Operator-learning architectures with shock-aware conditioning
\cite{peyvan2026,roohi2025fusion} and gradient- or zone-weighted losses
\cite{roohi2026mnf} exist precisely because vanilla architectures fail
here first; but loss weighting is a treatment of symptoms, and the
underlying tension---smooth function approximators versus
quasi-discontinuous targets---remains.

Taken together, these features mean that headline results imported from
smooth-PDE benchmarks (Burgers, lid-driven continuum cavities,
Darcy flow) transfer to rarefied gas dynamics only after substantial
qualification. They also mean, more positively, that rarefied gas
transport is an excellent stress test: an ML methodology that survives
honest evaluation here---noise-aware metrics, extrapolation splits,
constraint audits, perturbations of boundary conditions---has demonstrated
something real.
\section{What ``physics-informed'' guarantees---and what it cannot}
\label{sec:constraints}

The adjective ``physics-informed'' has become a quality signal, attached
to soft loss penalties and hard architectural constraints alike. The
distinction matters more in kinetic theory than almost anywhere else,
because the relevant physical structures---conservation, positivity of
$f$, realizability of moment sets, entropy dissipation, correct
asymptotic limits---are not decorative: violating them produces solvers
that crash or, worse, converge to plausible nonsense.

\subsection{Soft penalties versus hard structure}

A conservation term added to a loss function reduces the average
violation on the training distribution; it guarantees nothing pointwise,
nothing out of distribution, and nothing after the optimizer trades it
against the data term. Hard structure---convex network parameterizations
that preserve entropy-closure properties \cite{schotthofer2022},
relaxation architectures that inherit positivity, conservation, the
H-theorem, and the hydrodynamic limit by construction
\cite{xiao2023relaxnet}, asymptotic-preserving formulations whose loss
remains well-conditioned as $\mathrm{Kn}\to0$ \cite{jin2023apnn,%
jin2024apnn}, conservation projection of post-collision states in
learned-collision DSMC \cite{roohi2026collision}, and constraint terms
tied to measurable transport coefficients \cite{roohi2026kernels}---costs
expressivity and engineering effort, and buys properties that hold
unconditionally. My recommendation is blunt: every paper should declare,
for each claimed physical property, whether it is (H) hard by
construction, (S) soft-penalized, or (A) merely audited a posteriori; and
every paper should publish the a posteriori audit regardless---mass,
momentum, and energy balances over control volumes, minimum values of
$f$, entropy-production sign statistics. In our experience the audit is
cheap and often revealing, which is precisely the argument for
making it standard.

\subsection{Loss design as physics modeling}

Between soft and hard lies a productive middle ground: loss functions
engineered from problem structure. Gradient- and shock-sensitive
weighting concentrates capacity where smooth approximators fail
\cite{peyvan2026,roohi2025fusion}; far-field and monotonicity terms
suppress spurious oscillations in expansion regions; zonal losses assign
different physics penalties to physically different regions of a
separated rarefied flow \cite{roohi2026mnf}. These devices demonstrably
help. They also introduce hyperparameters whose tuning constitutes
undeclared model selection on the test problem---a quiet form of
leakage---and they can mask, rather than fix, an architecture's
inability to represent the target. The discipline that keeps loss
engineering honest is ablation reporting (results with and without each
term, on extrapolative splits) and sensitivity reporting (results across
a stated range of weights), neither of which is yet customary.

\subsection{Identifiability: macroscopic data underdetermine kinetic
states}

The deepest limitation is not architectural but informational. Most ML
models in this field are trained, and almost all are evaluated, on
low-order macroscopic fields. But the map from distribution functions to
low-order moments is many-to-one, and the practical question is how badly
this non-uniqueness contaminates conclusions. Two recent results from my
group quantify the problem in canonical settings. In physics-informed
reconstruction of Bhatnagar--Gross--Krook normal shocks, the
high-velocity tails of $f$---which control precisely the
non-equilibrium content---are only weakly observable from bulk profiles:
materially different tail states are compatible with macroscopic data at
the level of realistic noise, so confident tail recovery claims are
artifacts of the prior implicit in the network \cite{roohi2026tail}. In
the lid-driven rarefied cavity, reproducing the celebrated anti-Fourier
(cold-to-hot) heat flux \cite{akhlaghi2018srep} is sometimes presented
as a certificate of kinetic fidelity; we show it is not---models with
demonstrably incorrect fourth-order moment states reproduce the
anti-Fourier signature, so the signature cannot certify the closure
state \cite{roohi2026antifourier}.

The implications are general. (a)~Inverse problems---inferring
accommodation coefficients, closure relations, or boundary heat fluxes
from sparse macroscopic or gauge data---are ill-posed beyond a
problem-dependent information limit; ML does not repeal this, it merely
hides the regularization inside the architecture. Such studies should
report an identifiability analysis (even a local sensitivity/Fisher
analysis) before reporting point estimates. (b)~``Validation'' of an ML
kinetic model against macroscopic fields is weaker evidence than it
appears; agreement on $\rho$, $u$, $T$ is compatible with disagreement
on the distribution-level quantities (stress closure, heat-flux closure,
tails) that the kinetic description exists to provide. Where
distribution-level ground truth is available---and in simulation it
always is---papers should test against it directly. (c)~Conversely, when
only macroscopic agreement is achievable, claims should be scoped to
macroscopic quantities, only.
\section{Experimental verification: the missing leg}
\label{sec:experiment}

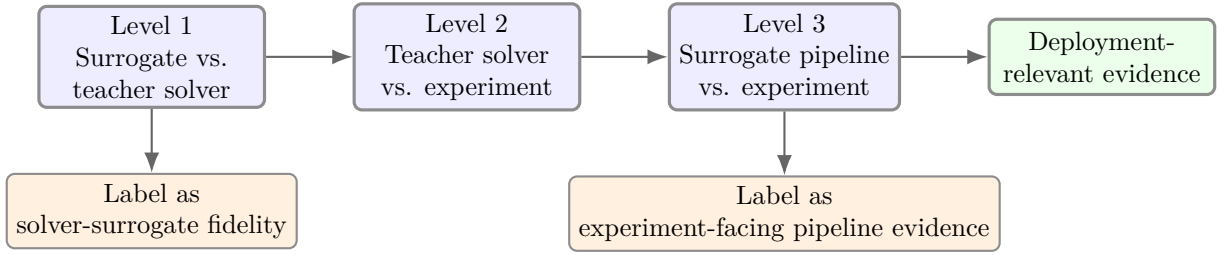
\begin{figure}[!tp]
\centering
\resizebox{0.97\textwidth}{!}{%
\begin{tikzpicture}[
    font=\small,
    node distance=1.15cm and 1.15cm,
    box/.style={rounded corners=3pt, draw=black!45, very thick, align=center,
        minimum width=3.0cm, minimum height=1.0cm, fill=blue!7},
    boxb/.style={rounded corners=3pt, draw=black!45, very thick, align=center,
        minimum width=3.0cm, minimum height=1.0cm, fill=green!8},
    claim/.style={rounded corners=3pt, draw=black!45, thick, align=center,
        minimum width=3.3cm, minimum height=0.85cm, fill=orange!12},
    arrow/.style={-{Latex[length=2.7mm]}, thick, draw=black!60}
]
\node[box] (l1) {Level 1\\Surrogate vs.\\teacher solver};
\node[box, right=of l1] (l2) {Level 2\\Teacher solver\\vs. experiment};
\node[box, right=of l2] (l3) {Level 3\\Surrogate pipeline\\vs. experiment};
\node[boxb, right=of l3] (deploy) {Deployment-\\relevant evidence};
\node[claim, below=0.85cm of l1] (claim1) {Label as\\solver-surrogate fidelity};
\node[claim, below=0.85cm of l3] (claim3) {Label as\\experiment-facing pipeline evidence};
\draw[arrow] (l1) -- (l2);
\draw[arrow] (l2) -- (l3);
\draw[arrow] (l3) -- (deploy);
\draw[arrow] (l1) -- (claim1);
\draw[arrow] (l3) -- (claim3);
\node[align=center, font=\footnotesize, above=0.55cm of l2, text width=8.5cm]
    {The claim should match the highest evidence level actually demonstrated.};
\end{tikzpicture}%
}
\caption{Three-level evidence hierarchy for ML models in rarefied gas
transport. Level~1 establishes fidelity to the training solver. Level~2
establishes that the teacher solver has an experimental anchor. Level~3
confronts the surrogate pipeline itself with experiment, ideally beyond
the training manifold.}
\label{fig:validation-ladder}
\end{figure}

A useful way to organize readiness for use is a three-level hierarchy.
Level~1 is \emph{verification against the training solver}: does the
surrogate reproduce the DSMC or Boltzmann-model solutions it was built
from? Level~2 is \emph{validation of the teacher solver against
experiment}: does the solver used to generate the training data reproduce
measured observables for the relevant class of flow? Level~3 is
\emph{direct confrontation of the surrogate pipeline with experiment},
ideally outside the training manifold. Most current studies in
rarefied-gas ML satisfy the first level; far fewer explicitly address the
third. That imbalance does not make the existing work unimportant, but it
should discipline the language of the claims: agreement with a teacher
solver is evidence of surrogate fidelity, not by itself evidence of
experimentally established physical fidelity. Figure~\ref{fig:validation-ladder}
summarizes the hierarchy.

\subsection{What experimental anchors exist}

The experimental record in rarefied gas dynamics is thin compared with
turbulence or aerodynamics, but it is not negligible, and it clusters
exactly where vacuum science needs it. Mass-flow-rate measurements
through microchannels span the hydrodynamic to near-free-molecular
regimes and resolve the Knudsen minimum \cite{ewart2007}; they remain the
sharpest integral test of slip, transition-regime transport, and GSI
combined. Thermal-transpiration and Knudsen-pump experiments, from
classical configurations to MEMS-scale multistage devices, provide
pressure-ratio and flow data in strongly thermally driven conditions
\cite{wang2020kp}. Radiometric-force experiments, including modern
re-examinations of the Crookes radiometer \cite{roohi2021crookes},
probe nonequilibrium edge and creep flows. Accommodation-coefficient
measurements---dedicated low-pressure heat-transfer assemblies
\cite{trott2011}, acoustic-resonator extraction \cite{sharipovmoldover2016},
and the compiled slip/jump database of \cite{sharipov2011data}---anchor
the boundary sub-model. In hypersonic rarefied conditions, the SR3
low-density wind-tunnel campaigns on the 70$^\circ$ blunted cone supply
density flow fields, aerodynamic forces, and surface heat transfer that
have served as the community's standard DSMC validation case for nearly
three decades \cite{allegre1997density,allegre1997heat}. And in vacuum
technology proper, pressure-profile predictions of test-particle Monte
Carlo codes are routinely compared with gauge readings in accelerator
systems \cite{kersevan2009}, while ML methods are already operational on
the measurement side, e.g.\ anomaly and heating detection from the LHC's
distributed vacuum-gauge network \cite{arpaia2021}.

\subsection{Why Level-3 verification is genuinely hard}

The obstacles to experimental verification are structural rather than
merely logistical. First, \emph{observability}: experiments in this field measure
integral, low-order quantities (mass flow, pressure ratio, force, heat
flux, line-of-sight densities), while ML models are trained and marketed
on full fields; by the identifiability arguments of
Section~\ref{sec:constraints}, agreement on integrals is a weak filter,
and disagreement is hard to attribute among GSI, geometry tolerance, and
model error. Second, \emph{confounded boundary conditions}: accommodation
is uncertain at the several-percent-to-tens-of-percent level depending on
surface state \cite{sharipov2011data,trott2011}, and in many benchmark
flows the predicted observable is at least as sensitive to accommodation
as to everything the ML model is supposed to capture; without independent
surface characterization, fitting and validating are indistinguishable.
Third, \emph{error-budget overlap}: experimental scatter, gauge
calibration drift, DSMC statistical noise, and surrogate error are
frequently the same order of magnitude, so a Level-3 comparison without a
quantified error budget on \emph{both} sides is uninterpretable. Fourth,
\emph{facility scarcity}: continuous low-density hypersonic facilities
are few, instrumented microflow rigs are bespoke, and the community has
no tradition of releasing digitized datasets with uncertainty
statements---papers report figures, not data. Fifth,
\emph{teacher-model ceiling}: a surrogate distilled from DSMC carries
DSMC's own model-form choices (collision model, internal-energy
relaxation, GSI); when it disagrees with experiment, the surrogate is
often not the primary source of error, while inherited physics may be
responsible. This makes Level-3 testing of the \emph{surrogate} per se
subtle. For applications, however, the relevant question is pipeline
validity rather than surrogate fidelity in isolation.

\subsection{A constructive program}

Four steps would change the situation within a few years, and none
requires new physics. (1)~\emph{Adopt paired benchmarks}: for each
canonical configuration with experimental data (microchannel mass flow
\cite{ewart2007}, Knudsen-pump stages \cite{wang2020kp}, the 70$^\circ$
cone \cite{allegre1997density,allegre1997heat}), curate a public package
of geometry, conditions, experimental values with uncertainties, and a
reference kinetic solution---so that every surrogate paper can run
Level-2 and Level-3 checks at near-zero cost, and a model trained on
simulation can at least be \emph{scored} against nature. (2)~\emph{Make
GSI part of the claim}: report which accommodation model and value
underlie the training data, propagate a stated accommodation uncertainty
through the surrogate, and present the experimental comparison as a band,
not a line. (3)~\emph{Use ML where it genuinely helps experiments}:
fast surrogates make Bayesian inference of accommodation coefficients
and outgassing parameters from mass-flow and gauge data computationally
routine; this inverts the relationship---instead of experiments
validating ML, ML extracts more physics from existing experiments, with
the identifiability caveats of Section~\ref{sec:constraints} stated up
front. The vacuum-systems community is well positioned here: distributed
gauge networks \cite{arpaia2021} plus fast learned forward models of
molecular-flow conductance amount to a digital-twin capability for
pump-down transients and leak/outgassing localization that test-particle
codes alone are too slow to support interactively \cite{kersevan2009}.
(4)~\emph{Label honestly}: a model verified only at Level~1 should be
called a solver surrogate, and its accuracy statements should name the
solver, not the physics.
\section{Open challenges and a reporting standard}
\label{sec:challenges}

\begin{figure}[!tp]
\centering
\resizebox{0.82\textwidth}{!}{%
\begin{tikzpicture}[
    font=\small,
    node distance=0.58cm,
    flowstep/.style={rounded corners=3pt, draw=black!45, very thick, align=center,
        minimum width=8.0cm, minimum height=0.82cm, fill=gray!8},
    focus/.style={rounded corners=3pt, draw=black!45, very thick, align=center,
        minimum width=8.0cm, minimum height=0.82cm, fill=blue!7},
    callout/.style={rounded corners=3pt, draw=black!40, thick, align=center,
        minimum width=2.6cm, minimum height=0.65cm, fill=orange!12, font=\footnotesize},
    arrow/.style={-{Latex[length=2.4mm]}, thick, draw=black!60}
]
\node[focus] (scope) {Define task, deployment envelope, and claim scope};
\node[flowstep, below=of scope] (data) {Generate solver data with stated collision, chemistry, and GSI models};
\node[flowstep, below=of data] (noise) {Estimate DSMC noise floor and relevant experimental uncertainty};
\node[flowstep, below=of noise] (splits) {Design splits: interpolation, extrapolation, geometry/transfer};
\node[flowstep, below=of splits] (train) {Train model with declared hard, soft, and audited constraints};
\node[flowstep, below=of train] (audit) {Audit balances, positivity/realizability, UQ calibration, and cost};
\node[focus, below=of audit] (label) {Assign evidence label: Level 1, Level 2, or Level 3};
\foreach \a/\b in {scope/data,data/noise,noise/splits,splits/train,train/audit,audit/label}
    \draw[arrow] (\a) -- (\b);
\node[callout, right=0.85cm of noise] (rep) {replicate\\solver runs};
\node[callout, right=0.85cm of splits] (ood) {held-out\\regimes};
\node[callout, right=0.85cm of audit] (cost) {break-even\\queries};
\draw[arrow] (rep.west) -- (noise.east);
\draw[arrow] (ood.west) -- (splits.east);
\draw[arrow] (cost.west) -- (audit.east);
\end{tikzpicture}%
}
\caption{A reporting workflow for rarefied-gas ML studies. The audit is
not an afterthought: noise estimates, extrapolation splits, constraint
checks, uncertainty calibration, and cost accounting should be planned
before the deployment claim is written.}
\label{fig:audit-pipeline}
\end{figure}
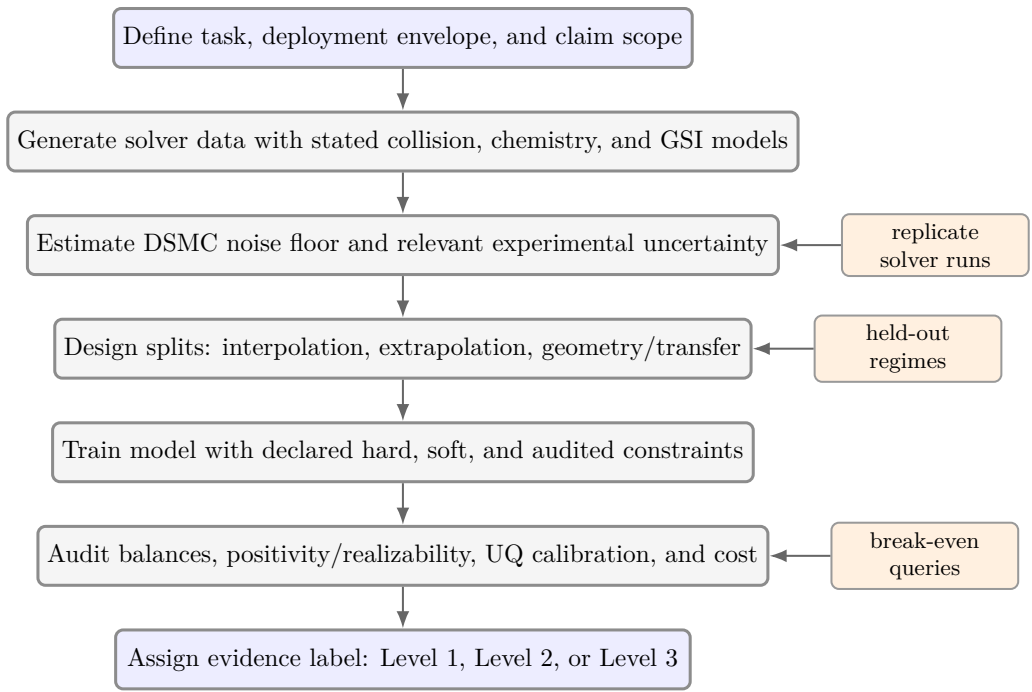

This section consolidates the recurring failure modes into explicit
challenges, and condenses the remedies into a checklist
(Table~\ref{tab:checklist}) that I would be content to see applied to my
own group's submissions. Figure~\ref{fig:audit-pipeline} summarizes the
recommended audit workflow.

\subsection{Generalization is claimed, interpolation is tested}

The single most common evaluation flaw is the random train/test split
over a dense, smooth parameter sweep. Adjacent DSMC cases differ by
small parameter increments and are strongly correlated; random splitting
therefore tests reconstruction between near-duplicates. The motivating
use cases---design exploration, optimization, off-nominal
prediction---are extrapolative or at least boundary-of-distribution. The
fix costs nothing: report, separately, (a)~interior interpolation error,
(b)~parameter-extrapolation error (hold out the extreme values of each
parameter), and (c)~where applicable, geometry- or
configuration-transfer error. Where we have applied this discipline to
our own surrogates, the gap between (a) and (b) is routinely a factor of
several---information the reader is entitled to.

\subsection{The noise floor defines honest accuracy}

For DSMC-trained models, the statistical error of the training fields
\cite{hadjiconstantinou2003} should be reported alongside every accuracy
metric. Errors below the estimated label-noise level should trigger a
replicate-based audit before being interpreted as physical accuracy. The
constructive corollary is noise-aware training (averaging ensembles of
independent DSMC runs, variance-aware losses) and noise-aware evaluation
(comparing surrogate-to-DSMC discrepancy against DSMC-to-DSMC rerun
discrepancy). For low-signal micro flows this is not pedantry; it decides
whether the reported percent-level errors mean anything at all.

\subsection{Cost accounting and the break-even query count}

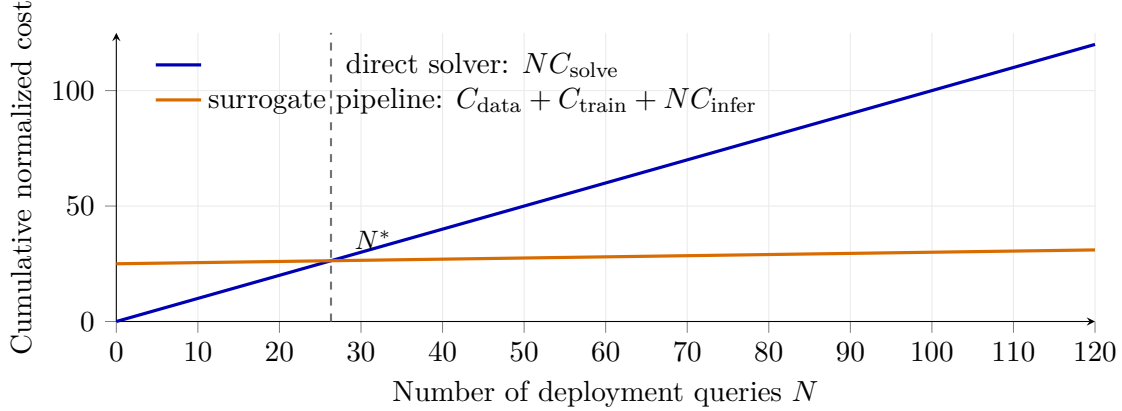
\begin{figure}[!tp]
\centering
\begin{tikzpicture}
\begin{axis}[
    width=0.88\textwidth,
    height=5.4cm,
    xmin=0, xmax=120,
    ymin=0, ymax=125,
    xlabel={Number of deployment queries $N$},
    ylabel={Cumulative normalized cost},
    axis lines=left,
    grid=both,
    grid style={black!8},
    legend style={draw=none, fill=none, at={(0.03,0.97)}, anchor=north west},
    tick align=outside,
    clip=false
]
\addplot[very thick, blue!70!black, domain=0:120] {x};
\addlegendentry{direct solver: $N C_{\mathrm{solve}}$}
\addplot[very thick, orange!85!black, domain=0:120] {25 + 0.05*x};
\addlegendentry{surrogate pipeline: $C_{\mathrm{data}}+C_{\mathrm{train}}+N C_{\mathrm{infer}}$}
\addplot[dashed, thick, black!55] coordinates {(26.32,0) (26.32,125)};
\node[anchor=south west, font=\small] at (axis cs:28,27) {$N^\ast$};
\end{axis}
\end{tikzpicture}
\caption{Break-even accounting for a surrogate. Per-query acceleration is
not the same as end-to-end acceleration; the offline data-generation and
training costs must be amortized over enough deployment queries. For the
simple one-fidelity model shown, $N^\ast=(C_{\mathrm{data}}+C_{\mathrm{train}})/(C_{\mathrm{solve}}-C_{\mathrm{infer}})$ when
$C_{\mathrm{solve}}>C_{\mathrm{infer}}$.}
\label{fig:breakeven}
\end{figure}

Speed-up claims should be end-to-end: data-generation cost (CPU/GPU-hours,
hardware), training cost, inference cost, and the resulting break-even
number of queries $N^\ast$ beyond which the surrogate pipeline is cheaper
than direct simulation. For one-off analyses $N^\ast$ is rarely reached;
for optimization, UQ, and digital-twin operation it often is. Stating
$N^\ast$ lets readers map a method onto their own use case, and its
absence converts an engineering trade into overstatement. Hardware
asymmetry (GPU surrogate vs.\ CPU solver) should be normalized or at
least disclosed, since several reported orders of magnitude reside in
the silicon, not the mathematics. Figure~\ref{fig:breakeven} makes the
accounting explicit.

\subsection{Transferability of boundaries, geometry, and gas models}

A surrogate is bound to the sub-models of its teacher: accommodation
coefficients, internal-energy relaxation numbers, collision potentials.
Papers should state these as part of the model card, and---where the
application warrants---demonstrate sensitivity by retraining or
fine-tuning across a plausible sub-model range. The GSI dependence is
the most consequential for vacuum applications and the least often
acknowledged (Sections~\ref{sec:why} and~\ref{sec:experiment}).

\subsection{Reproducibility}

Stochastic data, stochastic training: at minimum, seeds, splits, full
hyperparameters, and the training corpus (or its generator scripts)
should be released. Variance across training seeds should be reported
whenever a headline metric is within a factor of a few of that variance.
The field's credibility will be set by its worst accepted papers, not
its best.

\begin{table*}[!tp]
\centering
\caption{A minimal reporting checklist for ML studies in rarefied gas
transport. ``H/S/A'' = hard-by-construction / soft-penalized /
audited a posteriori.}
\label{tab:checklist}
\small
\setlength{\tabcolsep}{4pt}
\begin{tabular}{>{\raggedright\arraybackslash}p{3.4cm}>{\raggedright\arraybackslash}p{9.7cm}}
\toprule
Item & Required content \\
\midrule
Data provenance & Solver, collision and GSI sub-models, parameter box,
number of runs, per-field statistical noise level
\cite{hadjiconstantinou2003} \\
Split protocol & Interpolation \emph{and} parameter-extrapolation splits;
correlation structure of the sweep acknowledged \\
Physical constraints & For each property (conservation, positivity,
realizability, entropy, asymptotic limits): H, S, or A---plus the
a posteriori audit numbers in all cases \\
Baselines & Comparison against the strongest cheap baseline (linear
ROM/POD, Gaussian-process or RBF regression), not only against ``no
surrogate'' \\
Uncertainty & Method (ensembles, dropout, Bayesian) and an
out-of-distribution calibration check \cite{gal2016,psaros2023} \\
Cost accounting & Offline data + training cost, online cost, hardware,
break-even query count $N^\ast$ \\
Experimental anchor & Level-1/2/3 status declared
(Section~\ref{sec:experiment}); if Level-1 only, the model is labeled a
solver surrogate \\
Reproducibility & Code, seeds, splits, trained weights, data or
generators \\
\bottomrule
\end{tabular}
\end{table*}
\section{Outlook: a roadmap with falsifiable milestones}\label{sec:outlook}

Perspectives are obliged to predict, so I close with predictions that are
specific enough to be wrong, organized around four directions I consider
genuinely promising and one I consider premature unless it passes a clear transfer test.

\subsection{Structure-preserving surrogates as the default}
The clearest near-term win is the marriage of operator learning with hard
physical structure: architectures that conserve mass, momentum, and energy
by construction, guarantee positivity and moment realizability, and
dissipate an entropy functional, while retaining the parametric
flexibility of neural operators
\cite{schotthofer2022,xiao2023relaxnet,roohi2025enforced}. Soft penalties
were a reasonable first iteration; they should now be retired as the
primary mechanism for physical consistency. My expectation is that within
roughly five years, referees in this field will treat unconstrained
black-box surrogates of kinetic quantities the way they now treat
unvalidated turbulence models---publishable only when accompanied by explicit limitations and stronger
validation. If, by then, leading groups still publish soft-penalty-only
closures without realizability audits, this prediction will have failed
and the field will have a deeper incentive problem than I diagnose here.

\subsection{Active learning to place expensive kinetic runs}
The economics of Section~\ref{sec:challenges} change qualitatively when
the training set is not a fixed sweep but a budgeted sequence of DSMC or
deterministic-kinetic runs placed by an acquisition strategy that targets
predictive uncertainty. Because each DSMC run carries a tunable
statistical noise floor \cite{hadjiconstantinou2003}, rarefied gas
dynamics offers something most ML application domains lack: a dial that
trades sample cost against label noise. Multi-fidelity acquisition---many
noisy short-averaged runs to scout the parameter box, few long-averaged
runs to anchor it---is a natural fit and remains conspicuously
under-exploited. A concrete milestone: a published surrogate whose
training set was chosen by closed-loop acquisition, with a demonstrated
factor-of-several reduction in total CPU-hours to reach matched accuracy
against a uniform sweep, on a benchmark with declared splits.

\subsection{Vacuum science as the proving ground}
The vacuum community is, somewhat ironically, better positioned than the
aerospace community to demonstrate experimentally anchored ML. Free
molecular conductance calculations of the test-particle Monte Carlo type
\cite{kersevan2009} are fast per evaluation but are embedded in design
loops over geometry that multiply their cost; a geometry-conditioned
operator surrogate trained on Molflow-class output, validated against
measured conductances and pump-down curves of real assemblies, would be a
Level-2-to-Level-3 demonstration (in the sense of
Section~\ref{sec:experiment}) achievable with existing instrumentation.
Knudsen pumps \cite{wang2020kp} offer a second target: multi-stage
optimization is a genuine many-query problem, prototypes are bench-scale,
and mass-flow and pressure-head measurements are tractable. Accelerator
vacuum systems already log the operational data needed for ML-based
anomaly detection and inference \cite{arpaia2021}; extending those
pipelines from monitoring toward physics-grounded digital twins of gas
loads, with kinetic surrogates in the loop, is a realistic five-to-ten
year program with institutional owners (CERN-class laboratories) who can
sustain it.

\subsection{Hypersonics: inference before prediction}
For hypersonic non-equilibrium flows \cite{boydschwartzentruber2017}, I
expect the durable ML contribution in the next several years to be
inferential rather than predictive: Bayesian estimation of accommodation
coefficients, internal-energy relaxation parameters, and reaction-rate
uncertainties from sparse flight and wind-tunnel data, using fast
surrogates as likelihood engines. The identifiability results discussed in
Section~\ref{sec:constraints} are a warning, not a prohibition: they
delimit which parameters sparse macroscopic data can constrain, and
inference frameworks expose that limitation honestly as posterior width,
whereas point-predictive surrogates bury it.

\subsection{Foundation models: a falsifiable transfer test}
Large pre-trained models for PDEs or kinetic equations may eventually
prove useful in rarefied-gas applications, but transfer should be
demonstrated rather than assumed. The decisive comparison is not whether
a pre-trained model can be fine-tuned, but whether it outperforms a
specialist model under the same data and compute budget on held-out
geometries, gas--surface combinations, internal-energy or reactive
sub-model choices, and facility-specific boundary conditions. Those are
precisely the parts of rarefied gas transport where epistemic uncertainty
usually lives (Sections~\ref{sec:constraints} and~\ref{sec:experiment}).
Until such fixed-budget tests are reported, foundation-model language
should be framed as a research direction rather than as an established
solution strategy for vacuum or kinetic design workflows.
\section{Concluding remarks}

Machine learning will not replace kinetic solvers in rarefied gas
dynamics, and the interesting question was never whether it would. The
question is whether the field can convert a genuinely promising set of
tools---structure-preserving surrogates of collision physics, operator
models for many-query design, ML-assisted inference of boundary
parameters---into claims that survive contact with statistical noise,
honest data splits, full cost accounting, and, eventually, experiments.
The recurring failure modes documented in this Perspective (interpolation
reported as generalization, soft penalties reported as physical guarantees,
teacher-model agreement reported as physical accuracy, and sparse direct experimental verification) are not intrinsic to the methods.
They are reporting and incentive problems, and they are fixable by the
community that created them. The checklist of
Table~\ref{tab:checklist}, the verification hierarchy of
Section~\ref{sec:experiment}, and the falsifiable milestones of
Section~\ref{sec:outlook} are offered in that spirit. The vacuum and
micro/nano gas transport community, with its bench-scale experiments,
logged operational systems, and mature free-molecular codes, is uniquely
placed to supply what the ML-for-kinetics literature most lacks: an
experimental anchor. If this Perspective accelerates that supply, it will
have served its purpose.
\section*{Declaration of competing interest}
The author declares that he has no known competing financial interests or
personal relationships that could have appeared to influence the work
reported in this paper.

\section*{Data availability}
No new empirical data were created or analyzed in this Perspective. The
figures are conceptual schematics prepared for this article and are not
based on external datasets.

\section*{Declaration of generative AI and AI-assisted technologies in
the writing process}
During the preparation and revision of this work, the author used large
language-model assistants (Claude and ChatGPT) to support literature organization and language
editing. After using these tools, the author
reviewed and edited the content as needed and takes full responsibility
for the content of the published article.

\end{document}